# Controlled tempering of lipid concentration and microbubble shrinkage as a possible mechanism for fine-tuning microbubble size and shell properties

Intesar O. Zalloum,[acd] Amin Jafari Sojahrood, [acd] Ali A. Paknahad,[bcd] Michael C. Kolios,[acd] Scott S. H. Tsai,[bcde] Raffi Karshafian[acd]

a. Department of Physics, Toronto Metropolitan University, Toronto, Ontario M5B 2K3, Canada.

b. Department of Mechanical and Industrial Engineering, Toronto Metropolitan University, 350 Victoria Street, Toronto, Ontario M5B 2K3, Canada.

c. Institute for Biomedical Engineering, Science and Technology (iBEST), A Partnership Between Toronto Metropolitan University and St. Michael's Hospital, 209 Victoria Street, Toronto, Ontario M5B 1T8, Canada

d. Keenan Research Centre for Biomedical Science, Unity Health Toronto, 209 Victoria Street, Toronto, Ontario M5B 1W8, Canada

e. Graduate Program in Biomedical Engineering, Toronto Metropolitan University, 350 Victoria Street, Toronto, Ontario M5B 2K3, Canada

**ABSTRACT:** The acoustic response of microbubbles (MBs) depends on their resonance frequency, which is dependent on MB size and shell properties. Monodisperse MBs with tunable shell properties are thus desirable for optimizing and controlling MB behavior in acoustics applications. By utilizing a novel microfluidic method that uses lipid concentration to control MB shrinkage, we generate monodisperse MBs of four different initial diameters at three lipid concentrations (5.6, 10.0, and 16.0 mg/mL) in the aqueous phase. Following shrinkage, we measure MB resonance frequency and determine its shell stiffness and viscosity. The study demonstrates that we can generate monodisperse MBs of specific sizes and tunable shell properties by controlling MB initial diameter and aqueous phase lipid concentration. Our results indicate that the resonance frequency increases by 180-210% with increasing lipid concentration (from 5.6 to 16.0 mg/mL) while bubble diameter is kept constant. Additionally, we find that the resonance frequency decreases by 260-300% with increasing MB final diameter (from 5 to 12 μm), while lipid concentration is held constant. For example, our results depict that the resonance frequency increases by ~195% with increasing lipid concentration from 5.6 to 16.0 mg/mL, for ~11 μm final diameter MBs. Additionally, we find that the resonance frequency decreases by ~275% with increasing MB final diameter from 5 to 12 μm, when we use a lipid concentration of 5.6 mg/mL. We also determine that MB shell viscosity and stiffness increase with increasing lipid concentration and MB final diameter, and the level of change depends on the degree of shrinkage experienced by MB. Specifically, we find that by increasing the concentration of lipids from 5.6 to 16.0 mg/mL, the shell stiffness and viscosity of ~11 μm final diameter MBs increase by ~400% and ~200 %, respectively. This study demonstrates the feasibility of fine-tuning the MB acoustic response to ultrasound by tailoring MB initial diameter and lipid concentration.

## Introduction

Microbubbles (MBs), gas-filled shell-encapsulated agents from ~1–10 μm in diameter, are utilized in medical imaging and therapeutic applications.[1–9] The compressible core of MBs is typically composed of high-molecular-weight gases, that have low solubility.[2,3,5] The encapsulation shell is usually made of lipids, proteins, or polymers and is used to stabilize the bubble against rapid dissolution of the gas from the MB core.[4] In ultrasound imaging applications, MBs can be used to image blood perfusion and enhance blood-tissue contrast.[10–15] In therapeutic applications, MBs are used as targeted drug delivery vehicles whereby drugs are loaded on to the shells of MBs and are released when MBs are exposed to ultrasound through ultrasound-induced cavitation[16–22], sonoporation mediated applications (e.g. chemosensitization, and non-invasive disruption of the blood-brain barrier).[11,23–25] The optimization of these applications depends significantly on the response of the MBs to ultrasound.

The response of MBs to ultrasound depends on its resonance frequency, which depends on the MB size and shell physicochemical properties (composition and structure). Therefore, fine-tuning the size and shell characteristics of MBs is of great importance.[26–29] Conventional approaches to producing MBs, including agitation-based methods, produce MBs with relatively higher polydispersity compared to MBs generated using microfluidics. These polydisperse agents are, in principle, less desired because of their non-homogeneous response when exposed to ultrasound.[30] In addition to the lower polydispersity of MBs generated using microfluidic devices, microfluidics also allows for the generation of size-controlled bubbles. Nevertheless, control of the shell properties is still elusive. It has been suggested that the sensitivity of monodisperse MB populations is higher compared to polydisperse ultrasound contrast agents (UCA). The reason for this suggestion is that a small percentage of the MBs contribute to the overall echo when polydisperse agents are present.[27,28,30–34] Therefore, fine control of the size and the resulting resonance frequency of the MBs can result in a more effective MB response. However, in addition to its dependence on MB size, the resonance frequency is also greatly influenced by the shell properties of MBs. Currently, there is no widely adopted technique to create monodisperse MBs with tunable shell properties.

In this paper, we present a fundamental investigation of the influence of MB size and lipid concentration (and hence the amount of shrinkage experienced by MBs) on the resonance frequency and viscoelastic properties of MBs. In particular, we show the feasibility of fine-tuning the MB size, its viscoelastic properties, and as such, the resonance frequency of the MBs. To do this, we utilize a microfluidic approach that we recently developed, which uses lipid concentration to control MB shrinkage, to generate MBs of different final diameters that have gone through different degrees of shrinkage (before reaching the final stable equilibrium size).[35] We then compare the viscoelastic shell properties of the MBs that have undergone different amounts of shrinkage. We fit the linearized bubble model to the experimental frequency-dependent attenuation curves, and show, for the first time, that we can significantly change the shell viscosity and stiffness by changing the concentration of the lipids in the aqueous phase.

This study provides fundamental insights on fine-tuning the MB resonance frequency by tailoring the amount of shrinkage the MBs experience prior to reaching an equilibrium size. The information on MB resonant behavior are then used to, for the first time, quantify microfluidically generated MB shell parameters corresponding to different degrees of shrinkage. These results demonstrate the possibility of fine-tuning the acoustic response of MBs and may stimulate further interest in the investigation of MB membrane rheology. We first describe the methodology used to create MBs and control MB shrinkage. We then discuss the technique we use to assess shell parameters. Finally, we present and discuss the experimental attenuation measurement results and estimate the MB shell parameters.

## Materials and Methods

**Microfluidic device fabrication.** We use AutoCAD (Autodesk 2018, Inc., Dan Rafael, CA, USA) software to draw the microfluidic device design. Then, using a direct-write photolithography machine (MicroWriter ML3 Pro, Durham Magneto Optics Ltd, England), we construct our microfluidic device. This machine uses a 385 nm semiconductor light source and enables patterning of the microfluidic design directly onto the wafer, eliminating the mask aligner and photomask process.

Subsequently, we spin-coat a photoresist (SU-8 2025, MicroChem, USA) onto a silicon wafer. Then, we bake the silicon wafer at 65 °C for 3 minutes, and then again at 95 °C for 5 minutes. We use the MicroWriter ML3 printer to directly print our design onto the prebaked silicon wafer. Following this, we post bake the wafer at 65 °C for 3 minutes and at 95 °C for 5 minutes to stabilize the microfluidic device structure on the wafer. Finally, we remove the uncured photoresist using a developer solution. Profilometry shows that the microfluidic channels are 20 μm in height everywhere, the liquid and gas inlets are 60 μm in width, and the orifice is 8 μm in width and expands to a 360 μm wide in the downstream microchannel.

We pour a mixture of polydimethylsiloxane (PDMS, Sylgard 184, Dow Corning, USA) and curing agent in a ratio of 10:1 onto the silicon wafer mold. After that, we bake the mixture at 70 °C for two hours. We then remove the PDMS slab from the mold. Device inlets and an outlet are opened using a 1-mm diameter biopsy punch (Integra Miltex, Inc., Rietheim-Weilheim, Germany). To complete the microfluidic device (Fig. 1(a)), we bond the PDMS device to a glass slide by applying oxygen plasma (Harrick Plasma, Ithaca, NY, USA).

**Lipid formulation.** The lipid solution formulation used in our experiments is described elsewhere.[35] Briefly, we mix 100 mg of 1,2-dibehenoyl-sn-glycero-3- phosphocholine (DBPC), 1,2-dipalmitoyl-sn-glycero-3-phosphoethanolamine (DPPE), 1,2-dipalmitoyl-sn-glycero-3-phosphate (DPPA), and 1,2-distearoyl-sn-glycero3-phosphoethanolamine-N-[carboxy(polyethylene glycol)-2000] (DSPE-mPEG-2000) (Avanti Polar Lipids, Alabaster, AL, USA) in a mass ratio of 6: 2 :1 :1.[36] We heat 18 mL, 10 mL and 6.25 mL of phosphate-buffered saline (PBS) (Thermo Fisher Scientific Inc., Ontario, Canada) to 80 °C and add the heated PBS to the lipids to attain lipid solution concentrations of 5.6, 10.0, and 16.0 mg/mL, respectively. To ensure that all of the lipids are dissolved, we heat the solution at 80 °C for 15 minutes and sonicate at room temperature for another 15 minutes.

**Microbubble generation using microfluidics**

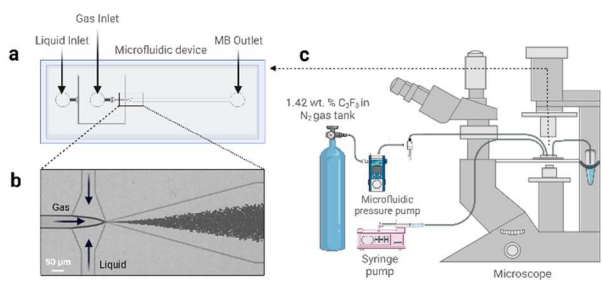

Figure 1. (a) Top view of the microfluidic device that we use in our experiments. The mixed gas phase and the aqueous lipid solution phase are forced towards the orifice, where the MBs are generated. Once the bubbles are generated, they travel through the 360 μm wide outlet channel and are then collected in a vial that is exposed to atmospheric pressure. (b) Image of the generation of MBs at the orifice of the microfluidic device. (c) The microfluidic experimental setup for producing the lipid-encapsulated bubbles using a flow-focusing design. The aqueous phase solution is comprised of lipid solution and the disperse phase is composed of a gas blend of 1.42 wt.% $C_3F_8$ in $N_2$. We use a syringe pump to carry the lipids into the aqueous phase inlet of the microfluidic flow-focusing device. We use a pressure pump to accurately deliver the gas from the gas tank into the disperse phase inlet of the microfluidic device. Diagram is not drawn to scale.

Fig. 1(a) shows the top view of the flow-focusing microfluidic device used in our experiments. The channel height for the microfluidic device is 20 μm and the orifice width where MBs are generated is 8 μm. By adjusting the gas pressure and liquid flowrate, we can control the initial size of the MBs at the site of generation.[37–39] Lower gas pressures and higher liquid flowrates decrease the diameter of the bubbles. A brightfield image for the formation of MBs using the flow-focusing device is shown in Fig.1(b).

A schematic representation of the experimental setup we use to generate MBs is shown in Fig. 1(c). The lipid solution is supplied to the liquid inlet of the microfluidic device using a syringe pump (Harvard Apparatus, PHD 2000, MA, USA). A blended gas composed of 1.42 wt.% octafluoropropane ($C_3F_8$) in nitrogen ($N_2$) (Messer Canada Inc., Mississauga, Canada) is injected into the gas inlet of the microfluidic device using a pressure pump (FLOW EZ, Fluigent, Paris, France), allowing for precise control and monitoring of the gas pressure at the gas inlet.

MBs are formed as the mixed gas disperse phase and the lipid solution continuous phase are forced through the orifice. By varying the continuous phase flowrate from 30 to 70 μl/min, and gas phase pressure from 8 to 18 psi, we generate bubbles with initial diameters of approximately 9 to 25 μm. We collect our MB samples in a vial that is exposed to atmospheric conditions.

To capture brightfield images of generated bubbles at the orifice, we use a high-speed camera (Phantom M110, Vision Research, NJ, USA) which is attached to an inverted microscope (AX10, Carl Zeiss AG, Oberkochen, Germany). We measure the initial diameter of the generated bubbles using ImageJ (National Institutes of Health, MD, USA) software. When the initial MB diameter is larger than the device channel height ( >20 μm), bubbles are "squeezed" into discoid shapes. When the initial MB diameter is larger than the device channel height ( >20 μm), we convert the diameter of the 'squeezed' MB to an equivalent spherical bubble.[38]

**Coulter counter measurements.** Using a Coulter counter (Beckman Coulter, ON, Canada), we measure the final MB size distributions. To ensure that MBs have reached their stable size, MB size measurements are performed 30 minutes after bubble collection.[35] Following that, attenuation experiments are conducted using MBs from the same vial. We report the mean, standard deviation, and polydispersity index (PDI) of the number-weighted MB sizes.[40]

**Attenuation measurements**

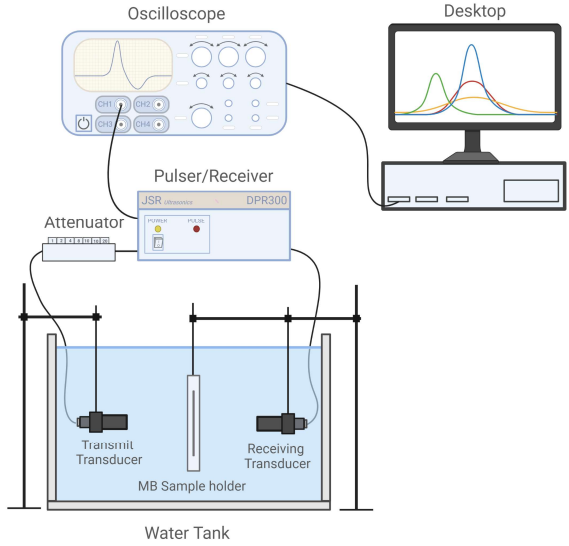

Figure 2. Schematic representation of the setup used to measure ultrasound attenuation. The ultrasound attenuation of the suspended MBs, placed into a 5 mm wide sample holder, is measured using the transmission and reception method. The diagram is not drawn to scale.

The ultrasound attenuation of the monodisperse lipid-coated MBs is measured through the transmission and reception method. This is widely accepted method for characterization of the shell properties of MBs.[41–48] The schematic representation of our experimental setup is shown in Fig. 2. In our experiments, we either use one pair of 2.25 MHz unfocused transducers (C304-SU, 2.25 MHz, 1" diameter, Olympus NDT, USA) or one pair of 1 MHz unfocused transducers (C302-SU, 1 MHz, 1" diameter, Olympus NDT, USA). Both transducers have 100% bandwidth. The transducers are aligned coaxially, facing each other, in a deionized water-filled tank. Microfluidically-generated lipid coated MBs were injected into the 5 mm wide sample holder, which was sealed using two acoustically transparent mylar sheets.

A pulser/receiver (DPR300, Imaginant Inc. Pittsford, USA) excites the transmit transducer that emits an ultrasound wave which travels through the MB suspension that is placed in the sample holder. An external 65-dB push button attenuator (50B-001 BNC, JFW Industries Inc., Indianapolis, IN, USA) is connected to the transmit transducer via BNC cable. The MB sample is interrogated using trains of broadband pulses with a pulse repetition frequency of 100Hz and averaged over 64 signals. By using a broadband transmission similar to what other researchers in the field have used,[8,43–47] the response of the MBs over a range of frequencies can be obtained in a short time. Electrical signals acquired by the receiving transducer are digitized (Agilent DSO-X 3024A; 12 bits, 4.00 GSa/s) and transferred to MATLAB software (MathWorks, Natick, MA) for offline analysis.

To compute experimental attenuation curves, signals are taken before and after placing the bubbles in the sample chamber. The measured attenuation coefficient (in dB/unit distance) is calculated as,[47]

$$\alpha_{measured}(f) = \frac{1}{l} 10 \log_{10} \frac{|V_{background}(f)|^2}{|V_{bubble}(f)|^2},$$

where $l$ is the total path length (5 mm) travelled by the ultrasound wave through the sample holder, and $V_{background}$ and $V_{bubble}$ are the frequency domain responses without and with MBs, respectively. Attenuation spectra measurements with each MB sample are repeated three times. We use a capsule hydrophone (HGL-0085 Hydrophone, ONDA corporation, CA, USA) to calibrate the acoustic pressure output. Peak negative pressure of 4.69 and 4.75 kPa are used in case of the 1 and 2.25-MHz transducers, respectively. An attenuator is used to minimize potential nonlinear effects,[27,49,50] by keeping the peak negative pressure sufficiently small. Low pressure sonication in linear regime has also been recently applied by Tabata et al.[46]

**Shell property estimates.** With the MB size distribution known, the MB shell stiffness ($S_p$ in N/m) and shell friction ($S_f$ in kg/s) are estimated by fitting a linearized encapsulated MB model proposed by de Jong et al.,[51,52] to the measured attenuation curves. The linearized model has been used extensively in previous literature studies;[41–47] it neglects multiple scattering effects and assumes low amplitude bubble oscillations (both of which are reasonable considering the high MB dilution ratios and low acoustic pressure employed in our experiments). For more information about the model, one can refer to previously published literature.[53,54] The estimated ultrasound attenuation per unit distance can be written as,

$$\alpha_{estimated}(r,f) = \frac{10}{\ln(10)} \sum_r n(r)\sigma_s(r,f) \frac{\delta_{tot}(r,f)}{\delta_{rad}(r,f)}.$$

Here, $n$ is the MB number density from measurements obtained using Coulter counter, $\sigma_s$ is the scattering cross sections of the MBs, and $\delta_{tot}$ is the total damping which is the sum of the radiation ($\delta_{rad}$), viscous ($\delta_{vis}$) and shell friction ($\delta_{sh}$) terms. Explicit equations for the total damping coefficient are available in Goertz et al.[55] Estimates of the shell stiffness and friction can be extracted by minimizing the error between the measured and estimated frequency-dependent attenuation coefficients.

## Results and Discussion

### Size distributions of stabilized shrunk MBs

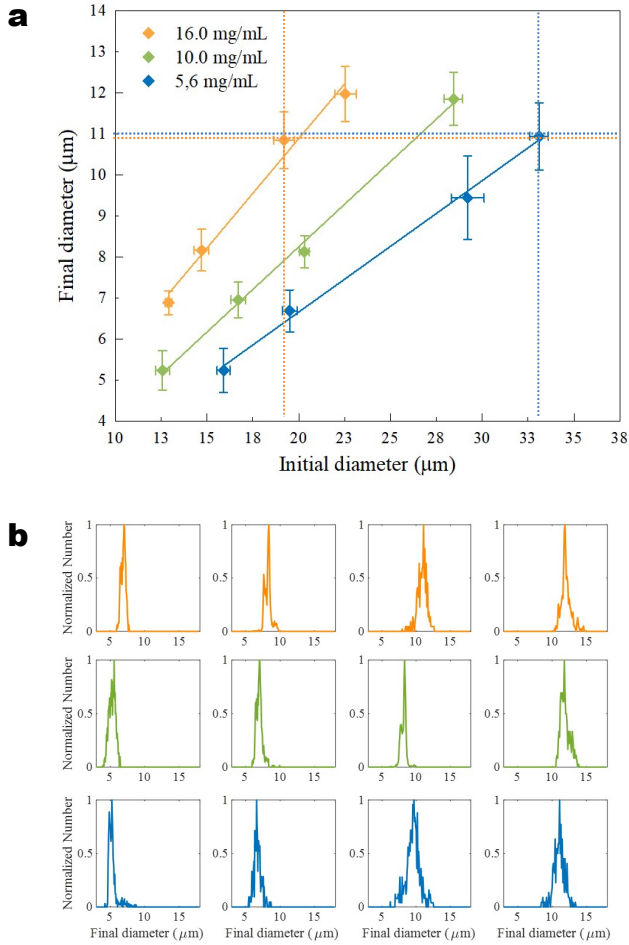

Figure 3. (a) Bubble final diameter versus initial diameter for lipid concentrations of 5.6, 10.0 and 16.0 mg/mL. The disperse phase gas is 1.42 wt.% $C_3F_8$ in $N_2$ for all experiments. Our results show that MB final diameter increases with increasing MB initial diameter for all lipid concentrations. Error bars represent standard deviations of the mean. (b) Normalized MB size distribution measurements obtained using Coulter counter using a 30 μm aperture. The first row (in orange), second row (in green) and third row (in blue) show the final MB diameters generated using lipid concentrations of 16.0 mg/mL, 10.0 mg/mL, and 5.6 mg/mL, respectively.

Traditional methods to produce MBs are mainly based on agitation and sonication.[56] These techniques are widely used in clinics because of their simplicity. However, these approaches generally produce polydisperse MBs, whose acoustic response is difficult to optimize. MB generation through microfluidics is advantageous because it allows for the production of size-controlled monodisperse bubbles.[27,29,37–39,57] A flow-focusing microfluidic device enables size-control of MBs at the orifice site.[30,58–63] The size of the microfluidically produced MBs can be varied by changing the inlet gas pressure, liquid flowrate, and the microchannel dimensions. The use of monodisperse MBs also provides a more defined attenuation peak compared to MBs that are polydisperse. This allows one to estimate the shell viscoelastic properties more accurately.

Using the flow-focusing microfluidic device, we generate lipid-coated MBs with final mean diameters ranging from 5 - 12 μm. Fig. 3(a) shows the bubble final diameters versus initial diameters for three different lipid concentrations: 5.6, 10.0 and 16.0 mg/mL. Our results show that, for a specific concentration of lipids, by increasing the initial bubble diameter $D_i$, one can obtain a larger bubble final diameter $D_f$. This suggests that one can obtain bubbles of specific final sizes by merely changing the initial bubble diameter. Our results also depict that by using different concentration of lipids, one can obtain bubbles with varying shrinkage levels, with more MB shrinkage obtained using lower lipid concentrations. These results are in agreement with recent findings reported by us.[35] These findings demonstrate that we can generate size-tunable monodisperse MBs, by simply changing the initial MB sizes or concentration of lipids used in our experiments. An average diameter fold reduction, $D_i/D_f$ of 3.1, 2.4, and 1.8 is found for bubbles generated using lipid concentrations of 5.6, 10.0 and 16.0 mg/mL, respectively. The first, second, and third rows in Fig. 3(b) show final MB diameters generated using lipid concentrations of 16.0, 10.0, and 5.6 mg/mL, respectively.

We find that the average PDI for the collected MBs is only 7.0 %. This value is significantly lower than reported PDI of up to 150% for MBs generated using sonication.[33]. Coulter counter measurements and attenuation experiments are conducted on MB samples from the same vial extraction.

**Attenuation measurements**

**A. Effect of lipid concentration on the resonance frequency**

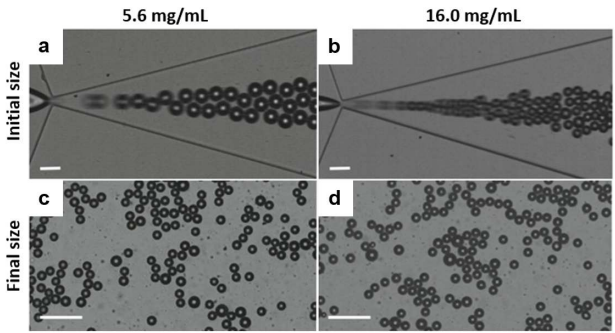

Figure 4. (a)-(b) Images of MBs at the generation orifice captured by a highspeed camera and (c)-(d) shrunk MBs after collection. We use lipid formulations with concentrations of 5.6 and 16.0 mg/mL to generate bubbles with an initial diameter, $D_i$, of approximately (a) 33 μm and (b) 19 μm. To capture the post-shrinkage diameters of the MBs, we transfer a few microliters of the collected MBs on to a glass slide. (c)-(d) In both cases, the mean final diameter, $D_f$, of the MBs is approximately 11 μm. Scale bars indicate 50 μm.

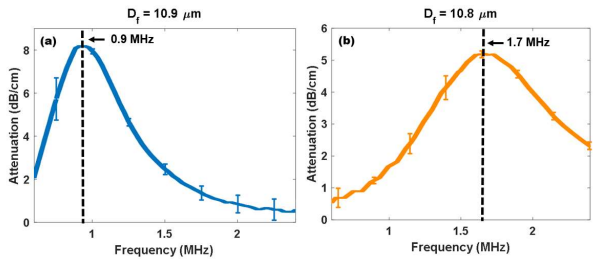

Figure 5. Attenuation coefficient of monodisperse MB samples of similar size at different lipid concentrations. (a) MB mean diameter of 10.9 μm generated using a 5.6 mg/mL lipid solution concentration, and (b) MB mean diameter of 10.8 μm generated using a 16.0 mg/mL lipid solution concentration. We determine

that, for a very similar MB size distribution, the resonance frequency is increased when MBs are generated with higher lipid concentrations in the aqueous phase.

To examine the effect of bubble shrinkage level (obtained by varying the lipid solution concentration) on the resonance frequency of the MBs, we generate MBs of similar final mean diameter $D_f$ using different lipid concentrations. The MBs generated using 5.6 and 16.0 mg/mL lipid concentrations are shown in Fig. 4(a) and 4(b), respectively. The mean initial diameter $D_i$ is ~33 μm (dotted blue vertical line in Fig. 3(a)) using the 5.6 mg/mL lipid solution, and the initial diameter $D_i$ is ~19 μm (dotted orange vertical line in Fig. 3(a)) for MBs generated using the 16.0 mg/mL lipid solution. In both cases, the mean final diameter, $D_f$, of the MBs is ~11 μm (shown by the dotted horizontal lines shown in orange and blue in Fig. 3(a)). Hence, the shrinkage level is decreased as the aqueous lipid concentration is increased. Specifically, we find that the diameter fold reduction $D_i/D_f$ is 3 and 1.8 for lipid concentrations of 5.6 and 16.0 mg/mL, respectively. Fig. 5 shows the attenuation coefficient for MBs of mean diameter ~11 μm generated using two different lipid concentrations. We find that MBs generated using the 5.6 mg/mL lipid concentration resonate at lower frequencies (~0.9 MHz), while MBs generated using the 16 mg/mL lipid concentration resonate at higher frequencies (~1.7 MHz). Next, we investigate the effect of the MB final diameter, $D_f$, on the frequency of attenuation peak, for MBs that have undergone the same shrinkage level.

**B. Effect of MB final diameter on the resonance frequency**

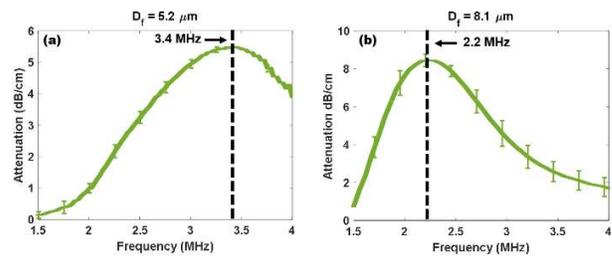

Figure 6. Frequency-dependent attenuation for mean final MB diameters, $D_f$, of (a) 5.2 μm and (b) 8.1 μm. In all cases, bubbles are generated using a lipid solution concentration of 10.0 mg/mL. The frequency of attenuation peak identified in the attenuation spectrum decreases with increasing MB size. Specifically, as the mean final MB diameter, $D_f$, increases from 5.2 to 8.1 μm, the frequency of attenuation peak decreases from 3.4 to 2.2 MHz, respectively.

To examine the effect of the mean MB final diameter, $D_f$, on the frequency of attenuation peak, we generate MBs with four diameters ranging from 5 to 12 μm at three lipid concentrations. Fig. 6 shows the attenuation experiments results' for the lipid-coated monodisperse MB suspensions of mean final diameter $D_f$ of (a) 5.2 and (b) 8.1 μm. In both experiments, the lipid solution concentration is 10 mg/mL. As expected, the frequency of the attenuation peak (resonance frequency) shifts to lower frequencies for larger mean MB diameters, as reported by Parrales et al. [31] and Doinikov et al. [64]. The resonance frequency is observed to decrease from 3.4 to 2.2 MHz with increasing final diameter, $D_f$, from 5.2 to 8.1 μm, respectively.

### C. Systematic control of bubble resonance frequency by using different lipid solution concentrations

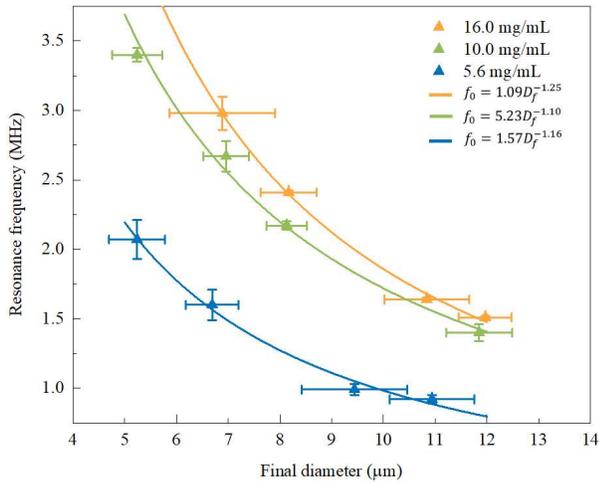

Figure 7. Resonance frequency versus MB final diameter for bubbles produced with a lipid solution concentration of 5.6 mg/mL (blue), 10.0 mg/mL (green), and 16.0 mg/mL (orange). Error bars represent standard deviations of the mean.

As stated above, we find that bubbles generated using different lipid solution concentrations show different levels of shrinkage. The bubble resonance frequency versus final bubble diameter for lipid solution concentrations of 5.6, 10.0, and 16.0 mg/mL is shown in Fig. 7. We use 1.42 wt.% $C_3F_8$ gas in all of our experiments. The resonance frequency decreases with increasing MB final diameter for all lipid concentrations. For example, the resonance frequency decreased by ~275% with increasing MB final diameter from approximately 5 to 12 μm for MBs generated using 5.6 mg/mL lipid solution concentration. We also determine that, for a very similar MB size distribution, the resonance frequency increases when MBs are generated with higher lipid concentrations. For example, our results depict that the resonance frequency increased by ~195% with increasing lipid solution concentration from 5.6 to 16.0 mg/mL, for ~11 μm final diameter MBs. Therefore, MBs of a specific resonance frequency can be achieved by changing the MBs' initial size and lipid solution concentration.

### Shell viscosity and shell stiffness versus bubble diameter

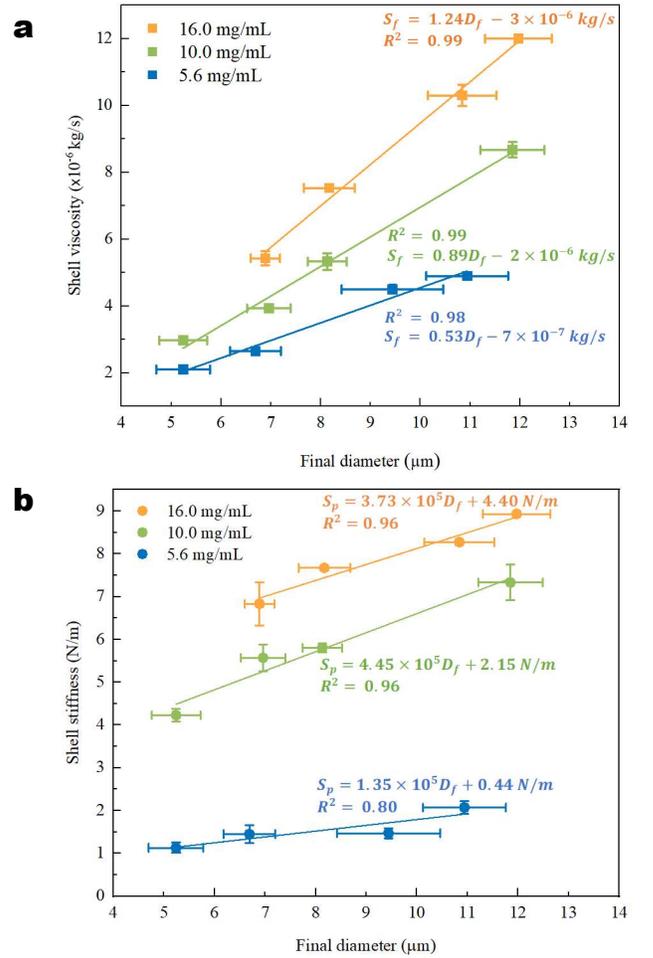

Figure 8. (a) Shell viscosity versus MB final diameter for bubbles generated using a lipid concentration of 5.6 mg/mL (blue), 10.0 mg/mL (green), and 16.0 mg/mL (orange). Shell viscosity linear regression: $0.53D_f - 7 \times 10^{-7}$, $0.89D_f - 2 \times 10^{-6}$, and $1.24D_f - 3 \times 10^{-6}\ kg/s$ for bubbles generated using lipid concentrations of 5.6, 10.0 and 16.0 mg/mL. (b) Shell stiffness ver-

sus MB final diameter for bubbles generated using a lipid concentration of 5.6 mg/mL (blue), 10.0 mg/mL (green) and 16.0 mg/mL (orange). Shell stiffness linear regression: $1.35 \times 10^5 D_f + 0.44$, $4.45 \times 10^5 D_f + 2.15$, and $3.73 \times 10^5 D_f + 4.40\ N/m$ for bubbles generated using lipid concentrations of 5.6, 10.0 and 16.0 mg/mL. Error bars represent standard deviations of the mean.

The behaviour of lipid-coated MBs when exposed to acoustic waves strongly depends on the MB size and shell parameters such as the shell stiffness and viscosity. Knowing the size distribution of these MBs, one can estimate the shell stiffness and viscosity from experimental acoustic attenuation measurements. Experimentally obtained attenuation versus frequency graphs are compared with linear model predictions to estimate the MB shell stiffness and viscosity. We observe in Fig. 8 that the shell viscosity and stiffness increase linearly with MB final diameter. Specifically, we find that the shell viscosity increases from $2.1 \times 10^{-6}$ to $4.9 \times 10^{-6}$ kg/s, $3.0 \times 10^{-6}$ to $8.7 \times 10^{-6}$ kg/s, and $5.4 \times 10^{-6}$ to $12.0 \times 10^{-6}$ kg/s with increasing MB mean diameters of 5.2 to 10.9 μm, 5.2 to 11.8 μm, and 6.9 to 12.0 μm, for lipid concentrations of 5.6, 10.0, and 16.0 mg/mL, respectively. Similarly, the shell stiffness increases from 1.1 to 2.1 N/m, 4.2 to 7.3 N/m, and 6.8 to 8.9 N/m with increasing MB mean diameters of 5.2 to 10.9 μm, 5.2 to 11.8 μm, and 6.9 to 12.0 μm, for lipid concentrations of 5.6, 10.0, and 16.0 mg/mL, respectively. These findings are in agreement with the experimental trends reported in previous studies.[42,65,66]

We plot the best-fit line for shell viscosity and stiffness versus bubble diameter at each lipid solution concentration and report the R-squared values in Fig. 8. For shell viscosity, we find that the R-squared values are 0.98, 0.99, and 0.99 at lipid solution concentrations of 5.6, 10.0, and 16.0 mg/mL, respectively. Additionally, we find the R-squared values for shell stiffness to be 0.80, 0.96, and 0.96 at lipid solution concentrations of 5.6, 10.0, and 16.0 mg/mL, respectively. These results suggest a linear association between viscoelastic shell parameters and final bubble diameter.

Our results depict that the estimated shell viscosity increases with MB diameter. This agrees with the findings of other researchers in the field. [42,67–74] Additionally, our results indicate that bubbles of the same final diameter, $D_f$, but generated using higher lipid concentrations have significantly higher shell viscosity and stiffness. For instance, we find that for ~11μm final diameter MBs, the shell viscosity and stiffness increase by approximately 200 % and 400 %, respectively, when we increase the lipid solution concentration from of 5.6 to 16.0 mg/mL. This suggests that, for bubbles of the same mean diameter, increasing the concentration of the lipids can modify the physical properties of the MB shell by increasing the shell stiffness and the viscosity.

The bubble behaviour is strongly influenced by the shell encapsulation.[47,75–77] Depending on the exposure parameters, adding a shell with a specific shell stiffness modifies the backscattered response and significantly increases the stability of the MBs.[75,76] The resonance frequency increases with increasing the shell stiffness.[75,76] Similarly, the viscosity of the shell reduces the MB expansion ratio and limits the collapse strength. The shell viscosity dissipates the acoustic energy by creating friction.[75] Sojahrood et al. show that the dissipation due to shell viscosity is the dominant dissipation mechanism leading to energy loss and increased damping. [78,79]

The effect of shell engineering of MBs for specific biological and biomedical applications has been studied. Sojahrood and Kolios, [80] and Cherin et al.[81] find that, for the same size MBs, increasing the shell stiffness increases the resonance frequency. This can have applications in enhancing the super-harmonic imaging by increasing the resonance frequency of the MBs over the applied frequency and thus amplifying the 2nd,3rd or 4th harmonic of the scattered signal by MBs.[81–83] Nonetheless, reducing the resonance frequency may have advantages in subharmonic imaging of MBs within the clinical frequency range.[84,85] This is because subharmonic emissions are more efficiently generated when the MB is sonicated at twice its resonance frequency.[86] Thus, reducing the resonance frequency of the MBs below the applied frequency will enhance the subharmonic generation. Moreover, reducing the shell viscosity enhances the generation of subharmonics at lower pressures and increases the signal amplitude. Finally, enhancing the shell viscosity has advantages in drug delivery applications and in opening the blood brain-barrier, that operate in the range of 300-500 kPa. In these applications a focused transducer applies a sharp pressure gradient at the focus. [34,79,87] Higher shell viscosity suppresses the oscillations of the MBs at lower pressure regions which reduces the pre-focal activity of MBs. This potentially may have implications in reducing the shielding effect (shadowing).[49,88] For a detailed discussion one can refer to [34,79,87,88].

## Conclusions

This study presents a fundamental investigation of the influence of MB size and lipid concentration on the MBs' resonance frequency and viscoelastic properties. Our data demonstrates the possibility of fine-tuning MB viscoelastic properties by modifying the amount of shrinkage the MBs experiences prior to reaching equilibrium diameter. In this study, we estimate the viscoelastic properties of lipid-coated MBs that have undergone different shrinkage levels. To do this, we use three lipid concentrations (5.6, 10.0, and 16.0 mg/mL) in the aqueous phase to generate monodisperse MBs of different initial diameters. We observe that MBs generated using the higher lipid concentration experience less shrinkage than those generated using lower lipid concentrations. We then acoustically characterize the MB suspensions using attenuation measurements to examine their shell properties, specifically the shell stiffness and viscosity. Our experimental results show that the shell viscosity and shell elasticity vary with the MB equilibrium diameter and lipid concentration.

## Funding Sources

R Karshafian (RGPIN-201505941), MC Kolios (RGPIN-2017-06496), and SSH Tsai (RGPIN-2019-04618) acknowledge funding support from the Natural Sciences and Engineering Research Council (NSERC) Discovery grants program. Equipment funding is from the Canada Foundation for Innovation (CFI, projects # 30994, 36687 & 36442), the Ontario Research Fund (ORF), and Toronto Metropolitan University.

## Notes

The authors declare no competing financial interest.

## ACKNOWLEDGMENT

The authors would like to thank Professor Stephen R. Euston (from Heriot-Watt University) and Professor Agata A. Exner (from Case Western Reserve University) for useful discussion.